# Tonal harmony and the topology of dynamical score networks

Marco Buongiorno Nardelli

*Department of Physics and Division of Composition Studies, University of North Texas, Denton, TX, 76203, USA*

*CEMI – Center for Experimental Music and Intermedia, University of North Texas, Denton, TX, 76203, USA.*

*iARTA – Initiative for Advanced Research in Technology and the Arts, University of North Texas, Denton, TX, 76203, USA.*

E-mail: mbn@unt.edu, websites: www.materialssoundmusic.com, www.musicntwrk.com
Tel. +1 (940) 369-8596

**Tonal harmony and the topology of dynamical score networks**

We introduce the concept of dynamical score networks for the representation and analysis of tonal compositions: a score is interpreted as a dynamical network where every chord is a node and each progression links successive chords. This network can be viewed as a time series of a non-stationary signal, and as such, it can be partitioned for the automatic identification of tonal regions using time series analysis and change point detection without relying on comparisons with pre-determined reference sets or extensive corpora. We demonstrate that the essential features of tonal harmony, centricity, referentiality, directedness and hierarchy, emerge naturally from the network topology and its scale-free properties. Finally, solving for the minimal length path through a route optimization algorithm on these graphs provides an abstraction of harmonic sequences that can be generalized for the conception of generative models of tonal compositional design.

*Keywords*: tonality, theory, music analysis, composition, music information retrieval.

## 1. Introduction

In the article *Topology of Networks in Generalized Musical Spaces*, recently published on the Leonardo Music Journal, (Buongiorno Nardelli, Topology of Networks in Generalized Musical Spaces, 2020) the author introduces the concept of harmony as a network representation of the musical structures built out of all possible combinatorial pitch class sets in any arbitrary temperament. Inspired by a long tradition of network representations of musical structures such as the circle of fifths (Heinichen, 1969), the Tonnentz (Euler, 1739), and recent works on the spiral array model of pitch space (Chew, 2014), the geometry of musical chords (Tymoczko D. , The Geometry of Musical Chords, 2006) and generalized voice-leading spaces (Callender, Quinn, & Tymoczko, 2008) (Tymoczko D. , The Generalized Tonnetz, 2012), the author interprets the harmonic structure of a composition as a large-scale complex network whose topological properties uncover its underlying organizational principles and demonstrates how classifications or rule-based frameworks (such as common-practice harmony, for instance) can be interpreted as emerging phenomena in a complex network system. Since the conclusions of that study serve as foundations for the present paper, we review some of its principal results.

Network analysis methods exploit the use of graphs or networks as convenient tools for modelling relations in large data sets. If the elements of a data set are thought of as "nodes", then the emergence of pairwise relations between them, "edges", yields a network representation of the underlying set. Similar to social networks, biological networks and other well-known real-world complex networks, entire dataset of musical structures can be treated as networks, where each individual musical entity (pitch class set (pcset), chord, rhythmic progression, etc.) is represented by a node, and a pair of nodes is connected by a link if the respective two objects exhibit a certain level of similarity according to a specified quantitative metric. Pairwise similarity relations between nodes are thus defined through the introduction of a measure of "distance" in the network: a "metric" (Albert & Barabási, 2002). As in more well-known social or biological networks, individual nodes are connected if they share a certain property or characteristic (i.e., in a social network people are connected according to their acquaintances, collaborations, common interests, etc.) Clearly, different properties of interest can determine whether a pair of nodes is connected; therefore, different networks connecting the same set of nodes can be generated.

In this paper we construct networks where nodes are the individual chords that can be extracted from the score, and edges are built between successive chords in the progression: nodes are connected if they appear as neighbours in the sequence. Naturally, nodes are visited numerous times, and the score evolution implies a directionality of the links. The networks are thus "directed", and each edge will have a weight (strength) proportional to the times the link is visited.

Given such network, we can perform many statistical operations that shed light on the internal structure of the data. In this work we will consider only two of such measures, degree centrality and modularity class (Barabasi & Posfai, 2016). The degree of a node is measured by the number of edges that depart from it. It is a local measure of the relative "importance" of a node in the network. Modularity is a measure of the strength of division of a network into communities: high modularity (above 0.6 in a scale from 0 to 1) corresponds to networks that have a clearly visible community structure. (Zinoviev, 2018). Isolating communities through modularity measures provides a way to operate within regions of higher similarity, and thus, as we will demonstrate in this paper, of closer harmonic content. Of particular interest in the context of this paper is the observation that tonal score networks exhibit scale-free properties, that is, their distribution of node degrees follows some form of power law, at least asymptotically. Scale-freeness is often associated with networks with few nodes with many connections (hubs) while the remainder exhibits relatively few connections (edges) (Barabasi & Posfai, 2016). The preferential attachment algorithm of Barabasi and Albert for the generation of scale-free networks (Barabasi & Albert, 1999), will be discussed in more detail in Section 6.

As summarized very effectively by the authors of (Moss, Neuwirth, Harasim, & Rohrmeier, 2019), tonal harmony compositions share four essential features: *centricity*, *referentiality*, *directedness*, and *hierarchy*. We argue here that all of these properties are completely captured by the network representation outlined above: *centricity*, the observation that tonal harmony is governed by a few central chords (Neuwirth, Harasim, Moss, & Rohrmeier, 2018), is captured by the degree distribution and the free-scale character of the score network, as discussed in (Buongiorno Nardelli, Topology of Networks in Generalized Musical Spaces, 2020) and in this paper. Chords do not occur in random order, but they are introduced following syntactical rules: *referentiality* introduces the idea that chords occur within a hierarchical structure at both global (among tonal regions) and local level (between individual chords) (Schoenberg, 1969) (Lerdahl & Jackendoff, 1983) (Rorhmeier, 2011) (Tymoczko D. , Root Motion, Function, Scale-degree: A Grammar for Elementary Tonal Harmony, 2003); *directedness*, is the preference for asymmetric chord progressions in tonal music (Hedges & Rohrmeier, 2011): it has direct implications in the formation of listening expectations and the directionality of chord progressions towards a build-up on expectation and release; finally, *hierarchy* involves the construction of hierarchical relationships at every level of the composition, from the locality of chords to the global organization of tonal regions and keys within a single piece (Schenker, 1954).

We now illustrate these key concepts and how they relate to the network representation of a given score with an example. In Figure 1, we show the score network of J.S. Bach's chorale from the Cantata "An Wasserfluessen Babylon", BWV 267. The edges here are directional, to indicate the pcsets progression in the piece (from now on, the term "pcset" and "chord" will be equally used to denote any vertical arrangement of pitch classes (Quinn, 2010)). A simple statistical analysis shows that the network has an average degree of 1.9 per node, and a modularity index of 0.5: the distribution of edges is relatively sparse, and many nodes are visited numerous times. One of the main result of such analysis is that the classification of nodes based on the modularity index clearly individuates tonal regions visited in the short piece (color-coded in Figure 1): G Major/C Major (blue), A minor (magenta), D minor (purple). Although we will come back for a more thorough analysis of the chorale in Section 5, we can still make some important observations: first, a statistical measure of modularity (a well-defined mathematical measure), individuates broadly the tonal regions of the piece without any *a priori* knowledge of the harmonic structure (*referentiality* and *hierarchy*); second, that the distribution of nodes' degrees gives us a way of identifying the individual tonal regions: the most important functional chords (tonic and dominant) are the ones that are visited more often (*centricity*); finally, the network is clearly directional: some progressions are more frequent than others, or *directedness*, a manifestation of the asymmetry of chord progressions in tonal harmony (Moss, Neuwirth, Harasim, & Rohrmeier, 2019).

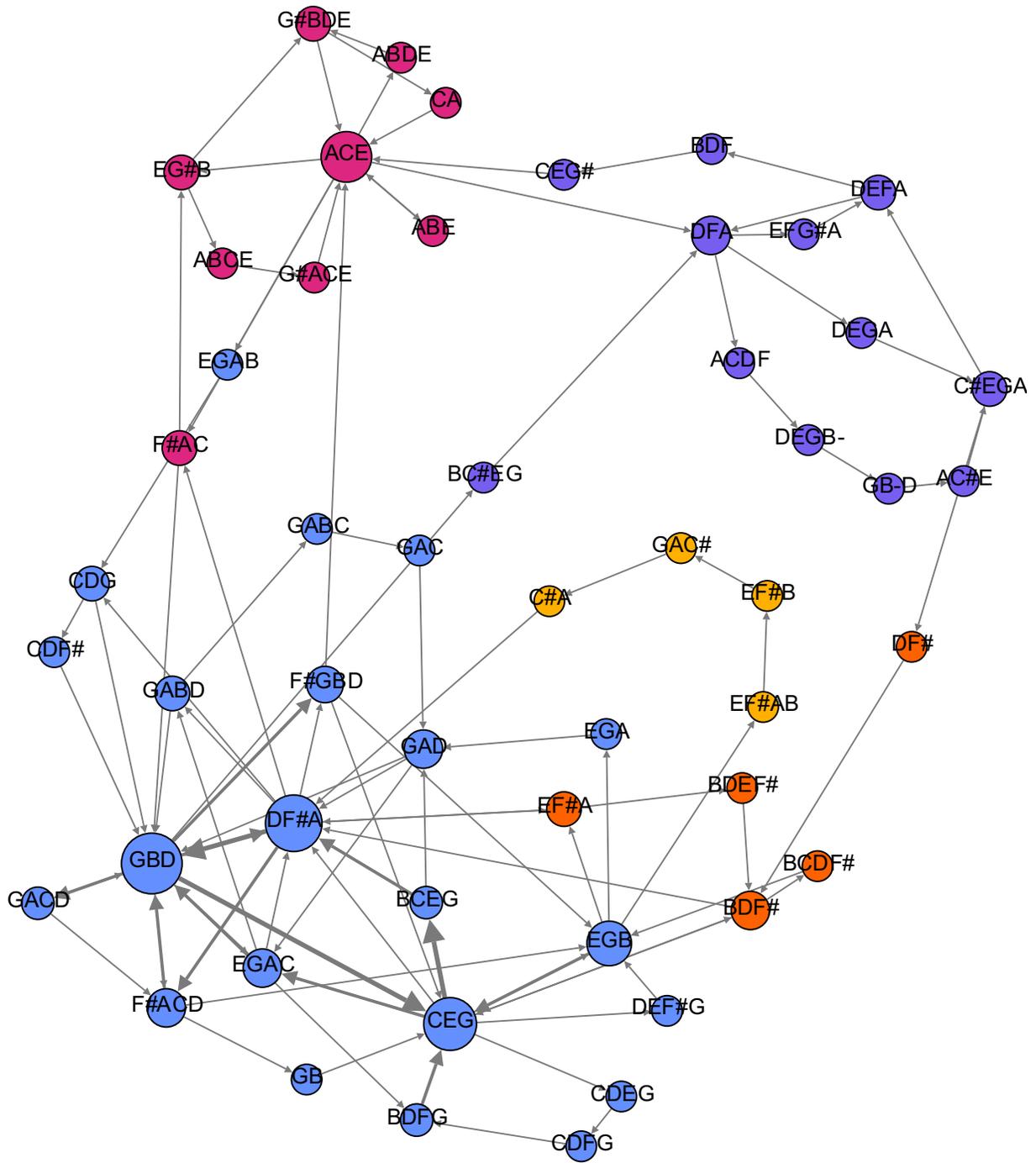

**Figure 1**. Score network of J.S. Bach's chorale from the Cantata "An Wasserfluessen Babylon", BWV 267. Nodes' radius is proportional to their degree and colours individuate different modularity classes. Edges are directional (arrows) to reflect the harmonic progression of the composition, and their thickness is proportional to the number of times that edge is visited. In this figure and in the following, network graphics and basic statistics are obtained using the Gephi app at www.gephi.org.

In the discussion above, we are taking a static perspective on the score network: we build its graph representation by collecting all the nodes (chords) and progressions (directional edges) from the whole piece. However, this is a procedure that ignores completely the fact that a piece of music is a temporal system, that is, each occurrence of any pcset is aligned according to a time scale, from the first to the last chord. This observation leads to the interpretation of the temporal network as a time series allowing the use of well-established techniques for time series analysis and change point detection in the evaluation of the harmonic content of the composition. In particular, we will demonstrate how we can automatically identify key regions in tonal compositions without relying on comparisons with pre-determined perceptual reference sets, such as in the Krumhansl-Schmuckler key-finding algorithm, (Temperley, 1999) or machine-learning of comprehensive corpora (Micchi, Gotham, & Giraud, 2020).

## 2. Methodology and data preparation

We will illustrate all the ideas and models using a particular example: the score of the string quartet in Eb Major, Op. 127 n. 12 by L. van Beethoven. However, all the algorithms and procedure developed here are generally applicable to any composition, even beyond those written in the tonal harmony idiom. We will make use of two principal software libraries for computational music analysis, both written in the Python language: MUSICNTWRK (at www.musicntwrk.com) and music21 (at https://web.mit.edu/music21). MUSICNTWRK is an open-source python library for pitch class set and rhythmic sequences classification and manipulation, the generation of networks in generalized music and sound spaces, deep learning algorithms for timbre recognition, and the sonification of arbitrary data (Buongiorno Nardelli, MUSICNTWRK: data tools for music theory, analysis and composition,, 2019). music21, developed at MIT (Cuthbert, 2010), is an object-oriented toolkit for analysing, searching, and transforming music in symbolic (score-based) forms of great versatility, whose modularity allows a seamless integration with MUSICNTWRK and other applications. All the results are compiled in a set of IPython Jupyter notebooks available from the author upon request.

Scores are read in musicxml format by the `readSCORE` function of MUSCNTWRK, where their harmonic content (and other relevant information, like in which bar the chord is found) is extracted using the `music21` parser and converter (using the "chordify" method). With this we obtain easily the full sequence of pcsets, chord by chord, where each change to a new pitch results in a new chord (Quinn, 2010). Upon "chordification", each pcset is reduced to its normal form. While such "quantization" of pcsets is quite adequate for the analysis of pieces with a harmonic movement where each vertical pcset plays some functional values (for instance in the corpus of J.S. Bach's chorales), for compositions where there is more contrapuntal development, the number of individual pcsets in the sequence can become very large, without providing necessarily more detailed information, since many such chords are only modifications via passing notes or vagrant harmonies. To make the analysis more manageable without losing any functional value, when appropriate we use a "filtering" algorithm based on the cumulative measure of how many times an individual pcset appears in the sequence (a more detailed discussion of the filtering procedure can be found in Section S1 in the Supplementary Online Material (SOM)). In Figure 2 we show the time-series of the first movement of the Beethoven string quartet Op. 127 after the filtering.

From the plot in Figure 2 we can draw already some observations:

a. Repetition: since the onset of the composition, we observe a number of pcsets that are repeated with higher frequency
b. Clustering: there are clear temporal clusters that appear during the time evolution
c. Incrementality: new pcset are continuously added throughout the duration of the piece: new harmonic material is continuously added until the end.

These observations have a musical counterpart that will become more apparent in the following Section. For now, it is sufficient to say that repeating pcsets are probably a manifestation of the concept of centrality, while clustering and incrementality are likely a manifestation of the referentiality and hierarchy properties discussed above.

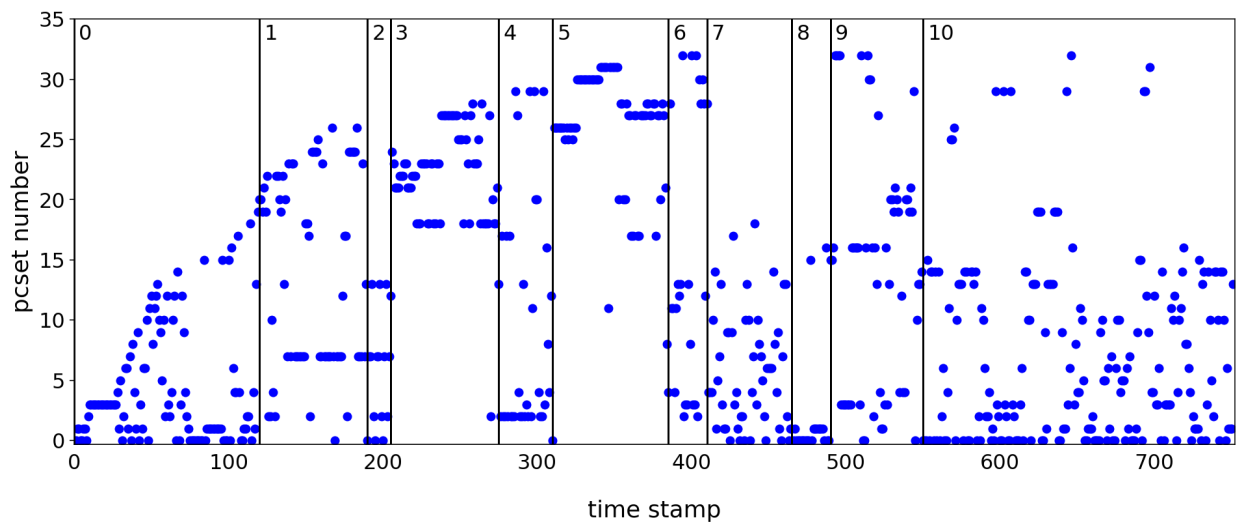

**Figure 2**. Time series of the filtered sequence of the normal ordered pcsets from the first movement of the string quartet Op. 127 n. 12 (blue dots). See Section 4 for the explanation of the regions identified by vertical lines.

## 3. The static score network

We can now build the score network as outlined in the Introduction: each of the unique pcsets are the nodes of a directional network, displayed in Figure 3. Nodes are labelled with the chord name; their radii are proportional to their degree (the number of connections each node has with the rest of the network) and are color-coded according to their modularity class. Similar to the case of Bach's chorale discussed above, also here belonging to a particular modularity class implies an underlining tonal region: we can broadly identify the magenta with the principal tonality, Eb major (the Eb major triad is central in this partition), C minor (yellow) and G minor (blue). However, a better understanding of the tonal dynamics of the piece can be gleaned only from the analysis of the changing topology during the time evolution of the network (see next Section).

**Figure 3**. Score network of the first movement of the Beethoven's string quartet Op. 127 n. 12.

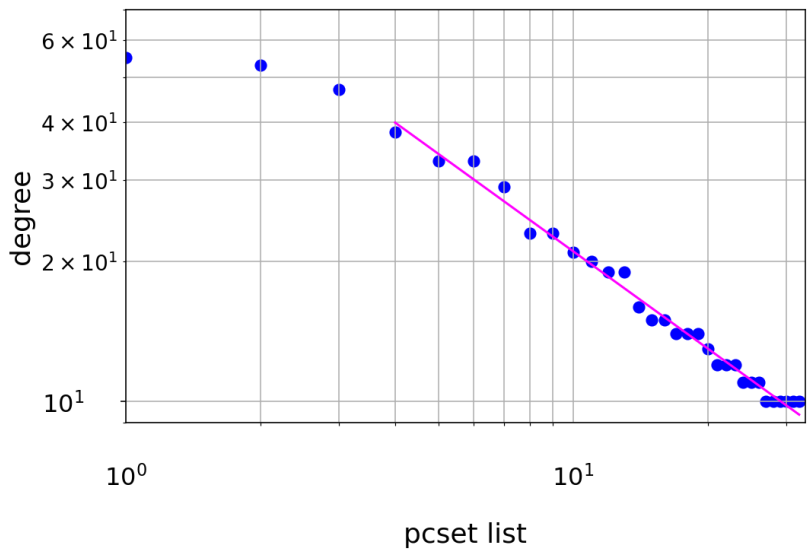

**Figure 4**. Loglog plot of the degree distribution in the score network. The magenta line is the best fit to a power law distribution $p(x) \propto x^{-\alpha}$ where the value of $\alpha = 2.53$ has been fitted using the `powerlaw` Python package (Alstott, Bullmore, & Plenz, 2014).

Here, we would like to extend our discussion with a quantitative analysis of the network topology and its scale-free properties. In Figure 4 we show, in a loglog scale, the degree distribution of the score network. As it is clear from the figure, there are few nodes with very high degree (centrality) that act as hubs, while the vast majority of them follows a power law distribution, a characteristic of scale-free networks (examples of power law distributions that might be familiar to the readers include the Pareto (80-20 rule) and the Zipf distributions). For a thorough discussion of the properties of scale-free networks and the implications for practical applications we refer the reader to (Barabasi & Posfai, 2016), here we summarize some of the most important consequences of scale-freedom. In the context of tonal harmony, the most important consequence of the appearance of hub chords in a score network is that their presence decreases the number of hops from one node to another: distances in a scale-free network are smaller than the distances observed in networks where the node degrees follow a random (flat) distribution. Hub chords act as "pivots" within progressions and between them, allowing for seamless transitions between tonal regions (they satisfy both the referentiality and hierarchy properties). After all, this is a manifestation of the elementary fact well known to any harmony student: the same chord coexists in different keys and with different functional properties, I in C major is also V in F major and so on. Our results align with the conclusions of a previous study of the statistical characteristics of tonal harmony in Beethoven string quartets, where the power law behaviour was inferred from the chord lexicon analysed as an *n*-gram model as in Natural Language Processing (Moss, Neuwirth, Harasim, & Rohrmeier, 2019). However, while Moss *et al.* relied on a manually annotated corpus (Neuwirth, Harasim, Moss, & Rohrmeier, 2018), our results stem from a purely statistical analysis of the raw pcsets material.

## 4. The dynamic score networks

The analysis conducted so far is only able to give us a broad perspective on the characteristics of the piece, without elucidating the hierarchy of the different regions in the overall compositional design. In order to arrive to this, we need to look at the score not as a static network, but as a dynamic graph: a time evolution where the system explores different regimes. A practical way to achieve this is to look at the pcset data in Figure 2 as a time series and exploit the analytical techniques that have been developed to tackle this kind of data systems. The observation of repetition, clustering and incrementality suggests that time-series data can indeed provide the information on the different regimes, the tonal regions. In this perspective, we analyse the pcset sequence of Figure 2 as a non-stationary signal and apply an algorithm of change point detection and segmentation that uses a cost function with a Gaussian kernel (Truong, Oudre, & Vayatis, 2020), as implemented in the `ruptures` Python package.

It is a sequential approach: first, one change point is detected in the complete input signal, then the series is split around this change point, and the operation is repeated on the two resulting sub-signals (Fryzlewicz, 2014). The cost function used in this work is:

$$c_{rbf}(y_{a \ldots b}) = (b-a) - \frac{1}{(b-a)} \sum_{s,t=a+1}^{b} \exp(-\gamma \parallel y_s - y_t \parallel^2)$$

where $y$ is a multivariate nonstationary signal that in the present work is the mapping of the individual pcsets to a sequence of integers (as shown in Figures 2), so that the signal to segment is one-dimensional, and (*b-a*) is the length of the sub signal $y_{a \ldots b}$.

The time series of the sequence of the normal ordered pcsets after the segmentation is partitioned, in Figure 2, by vertical lines that correspond to the breaking points. We identify 11 different regions (0 to10), where each segment corresponds to a chord range identified by the breaking points at the following bars: 32, 58, 63, 101, 120, 150, 166, 181, 187, 207. Just from a visual inspection, we can observe that each of the segments display characteristics that set it apart from the contiguous ones. With these results in hand, we can now proceed to the analysis of each individual region as a network to capture the topology of pcset interactions in a window of time. The result of this process is summarized in Figure 8, where we display the partitioning of the score as a multi-layered network, where each layer displays the connectivity of the pcsets in the range of bars identified above by the segmentation algorithm.

It is clear that each layer displays strikingly different topologies, where different groups of nodes are connected at different times, and the inter-layer links indicate the pcset that act as pivot chords between different tonal regions. For instance, in layer 0, that can be identified with the beginning section of the string quartet in Eb major (and indeed the Eb major triad has the highest degree here, hence is a hub in the power law sense) the G minor triad plays a very small role in the overall geometry of the layer network, but it becomes the most important node in the next layer, where Beethoven actually modulates to G minor. The vertical links between layers show the common nodes in the two networks and thus the connection between the folds of the scale planes: 0 expresses the chord structure of the Eb Major scale, and 1 the chord structure of the G minor scale.

**Figure 8**. Different sections of the score as identified by the segmentation algorithm displayed as a multi-layer network.

We can now apply the same analytical techniques that we considered for the full score network to these layer networks. The subdivisions of the score network now acquire a more definite musical meaning, reflecting the harmonic evolution of the compositional design. Before me move to the quantification of the harmonic content is important to make some observations that are at the same time structural and stylistic. Given the finite nature of the score, some subnetworks are comprised of very few nodes (see for instance layer 2 and 8), an indication of regions of the composition where there is less harmonic movement in preparation of more dramatic transitions. Of the more articulated sections, the ones that last longer and thus are comprised of more nodes and edges, still display a structure centred around a few main nodes (hubs with larger degree) and their degree distribution still largely follows a power law distribution that reflects their scale-free properties. This is evidence of the *referentiality* character of tonal harmony and of hierarchical structure: we can say that these properties manifest in the self-similarity characteristics of the tonal progressions.

The determination of the score sections and their representation as layer networks allows us to quantify the relationships between the different regions using well established algorithms for graph similarity measurements. Here we use a graph distance metric based on the maximal common subgraph, as originally proposed in (Bunke & Shearer, 1998) and implemented in the `GMatch4py` python library. The results of the comparisons between the layer networks of the sections are shown as a heat map in Figure 9.

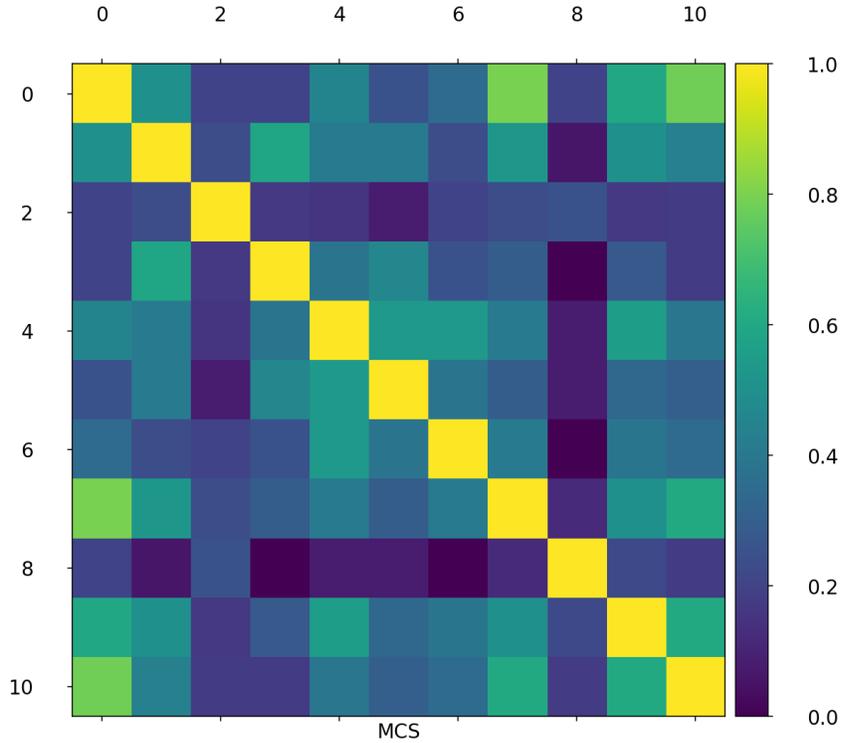

**Figure 9**. Heat map of the similarity measure between the layer networks.

Before we analyse these results more closely, we can relate the individual degree distribution of the layer networks to the attribution of a particular tonal character (key) of the individual regions. Given the predominance of the tonic and dominant in tonal music, we assume that the most frequently visited pcsets in each subnetwork are either major or minor triads or dominant seventh chords (see also (Pardo & Birmingham, 2002)). With this proviso, we obtain the assignment of the tonal regions shown in Table I. The partitions identified by our algorithm agree well with the expert analysis of the score from the ABC-1.0 data (Neuwirth, Harasim, Moss, & Rohrmeier, 2018), as it will be reviewed more in depth in the discussion of the full score analysis in Section 5 below.

| Section | measures | prevalent_chord | region |
|---|---|---|---|
| 0 | 1-32 | [Eb, G, Bb] | I |
| 1 | 32-58 | [G, Bb, D] | iii |
| 2 | 58-63 | [G, Bb, D] | iii |
| 3 | 63-101 | [G, B, D] | III |
| 4 | 101-120 | [C, Eb, G] | vi |
| 5 | 120-150 | [C, E, G] | VI |
| 6 | 150-166 | [Eb, G] | I |
| 7 | 166-181 | [Eb, G, Bb] | I |
| 8 | 181-187 | [Eb, G, Bb] | I |
| 9 | 187-207 | [G, Bb, Db, Eb] | IV |
| 10 | 207-282 | [Eb, G, Bb] | I |

**Table I**. Assignment of relative tonal regions in the first movement of the string quartet Op. 127 n. 12.

A comparison between the key region assignments and the similarity map for the sub-networks shows a greater variability than one might naively expect: even regions that have the same tonal character might show a relatively low similarity score. Notable exceptions are layers 0, 8 and 10 (the regions in the predominant key of Eb major) that have a similarity score of about 0.8 (they show an 80% overlap). Of course, stylistic considerations play a central role here, since the observed variability serves as an expressive tool in the development of the composition. Moreover, also the definition of "prevalent chord" will have to be adapted to the particular style or historical period. Finally, the determination of the tonal regions allows for the straightforward application of computational analysis tools, such as the ones in `music21`, for the identification of the chord types and their roman numeral representation (Tymoczko, Gotham, Cuthbert, & Aritza, 2019).

## 5. Score analysis, calibration and uncertainty quantification

So far, we have discussed the general principles behind our dynamical score network approach for harmonic analysis using the first movement of Beethoven's string quartet Op. 127, n. 12 without optimizing the two most important free parameters that enter into the algorithm. To summarize: first, in order to speed up the computational analysis we have the filtering of the chord sequence exemplified in Figure 3. This procedure allows for a reduction of the number of pcsets in the time series, and in general does not influence the outcome of the analysis if the threshold is kept lower than 10%: most of the rare pcsets arise from the usage of passing tones. Second, and most important, are the parameters that enter into the segmentation algorithm. In the situation in which the number of change points is unknown, the binary change point detection algorithm uses a penalty parameter (>0) that increases (smaller penalty) or reduces (larger penalty) the number of breaking points. Controlling the penalty value allows us to zoom from the harmonic macro- to micro-structure of the composition: if the cost of breaking the series is high (large penalty) we eventually obtain a single section that corresponds to the static network of Figure 5. From there we can basically extract the overall key of the composition (Eb major in mov. 1). Reducing the penalty, the number of breakpoints increases and so does the harmonic partitioning of the score. In all the examples so far, a penalty=3 was chosen.

In order to explore quantitatively the accuracy of the partitioning of the dynamical score network, we have compared the result of our analysis with the expert annotations of the Annotated Beethoven Corpus (ABC) (Neuwirth, Harasim, Moss, & Rohrmeier, 2018) for all the movement of the string quartet. In order to obtain a fair comparison between the manual annotations and the automatic partitioning, we have adjusted the penalty parameter in order to obtain a similar number of partitions in the two sets. The values chosen are thus: 1.8 for mov. 1, 2.8 for mov. 2, 2.6 for mov. 3, and 2.6 for mov. 4. On average, the agreement between the ABC and this work is greater than 80%, a remarkable result, considering a certain degree of arbitrariness in the assignment of roman numerals to chords in any composition: Mov. 1 match to 81%, mov. 2 to 89%, mov. 3 to 86% and mov. 4 to 77% (see Table S1 in SOM for the full analysis of the range of bars that corresponds to different harmonic regions). Moreover, we did not refine the definition of prevalent chord introduced for the automatic key identification, so further alterations can indeed improve upon the present numbers.

To conclude this section, we can revisit the chorale BWV 267 of Figure 1 and run the same algorithm. In general chorales are very short pieces with a small number of chords (BWV 267 contains only 17 bars and 123 chords of which 52 unique pcsets), a characteristic that makes them relatively poor candidates for analysis based on statistical inference. However, even in this case, our algorithms identify tonal regions with an accuracy of 85% relative to expert annotations by Jones, Tymoczko and Robb, as provided in the Bach chorale corpus of the `music21` package (Cuthbert, 2010) (Tymoczko, Gotham, Cuthbert, & Aritza, 2019) proving once more the accuracy and flexibility of the method.

## 6. Towards a generative model of tonal compositional design

The identification and classification of the dynamical score network as a convincing representation of a composition suggests avenues for the design of autonomous systems for the generation of harmonic progressions. Looking at any score network we can ask ourselves how many links we have to visit in order to complete a full path and travel all the chord progressions at least once. This is a well-known route optimization problem: to find a shortest closed path or circuit that visits every edge of a graph. When the graph has a Eulerian circuit (a closed walk that covers every edge once), that circuit is an optimal solution. Otherwise, the optimization problem is to find the smallest number of graph edges to duplicate (or the subset of edges with the minimum possible total weight) so that the resulting multigraph does have a Eulerian circuit (Edmonds & Johnson, 1973). If we analyse the network of Figure 1 in this context, Bach completes his circuit in 123 steps on a network of 52 nodes and 101 directed links. This means that in Bach's circuit 22 edges have been duplicated. If we run the route optimization algorithm on the same network, we find a minimum circuit length of 110. This minimal path overlaps with the Bach's version. Its shorter length is only due to repetitions of some of the loops in the original score, while the optimal Eulerian path does not allow for double counting. It is important to note that most loops in the graphs can be visited clockwise or counter clockwise without changing the total path length, thus generating multiple "minimal length paths". This allows to retrieve the analogy with the Bach's original straightforwardly (see Figure S3 in SOM for a direct comparison and a more detailed discussion). Since this solution is built imposing *a priori* the *directedness* of the underlying tonal harmony, the resulting circuit provides an abstraction of the harmonic evolution of the piece.

So far, this might seem just an interesting observation on a single chorale: how general is this result? In order to answer this question, we have run the route optimization problem on the full Bach corpus of `music21`, for a total of 433 scores. For each score we compared two distinct cases with Bach's original circuit: the Eulerian path built on the original score network, and a Eulerian path on a randomly generated scale free network with a pcset distribution that parallels the one of the original scores.

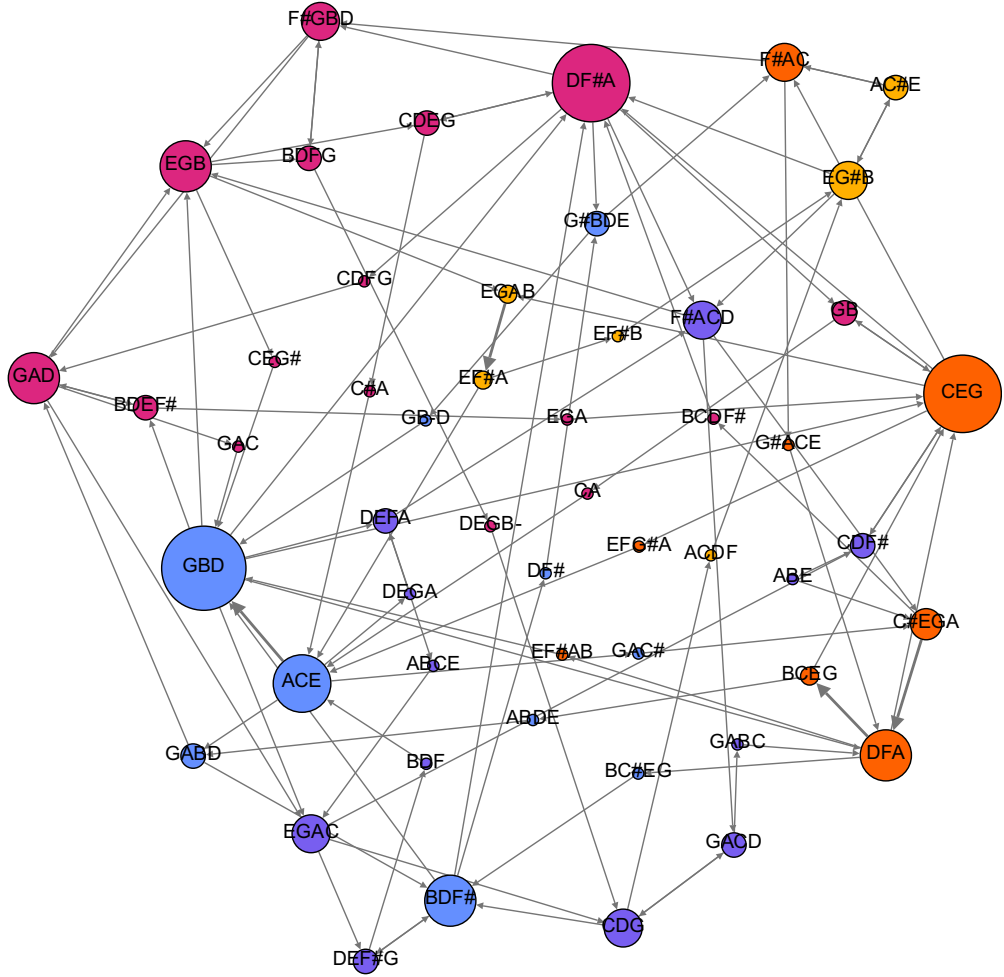

**Figure 9.** Scale-free network generated with the Barabasi-Albert algorithm including the chord-node assignments and one of the directional Eulerian paths.

There are many generators for scale-free networks (Hagberg, Schult, & Swart, 2008). In this study we use the original model proposed by Barabasi and Albert (Barabasi & Albert, 1999). It incorporates two important general concepts, growth and preferential attachment: growth means that the number of nodes in the network increases over time; preferential attachment means that the more connected a node is, the more likely it is to receive new links. Nodes with a higher degree have a stronger ability to grab links added to the network. Both characteristics ensure that the generated network displays scale-free properties and fall into the general taxonomy of tonal harmony, where centrality is a representation of preferential attachment and growth is intrinsic to the dynamical score network concept. It is then straightforward to generate random scale-free networks that can be interpreted as the skeleton of a dynamical score networks. As illustration, we show in Figure 9 a Barabasi-Albert network where we have assigned pcsets to nodes following the ranking of the degree distribution of BWV 267. This gives the following for the six highest degree nodes: GBD (degree 13 in BWV 267 and 13 in the Barabasi-Albert (BA)), DF#A (11 and 13, respectively), CEG (10 and 11), ACE (9 and 11), EGB (8 and 8), and DFA (8 and 6). Finally, we run the route optimization algorithm to generate a Eulerian path. From Figure 9 we immediately observe that this new network still maintains some of the characteristics of the original: nodes are distributed in modularity classes that do not deviate much from the original and indicate a division in tonal regions that carries over from the chord distribution. A close inspection of the path also shows the permanence of common progressions inherited from the original network.

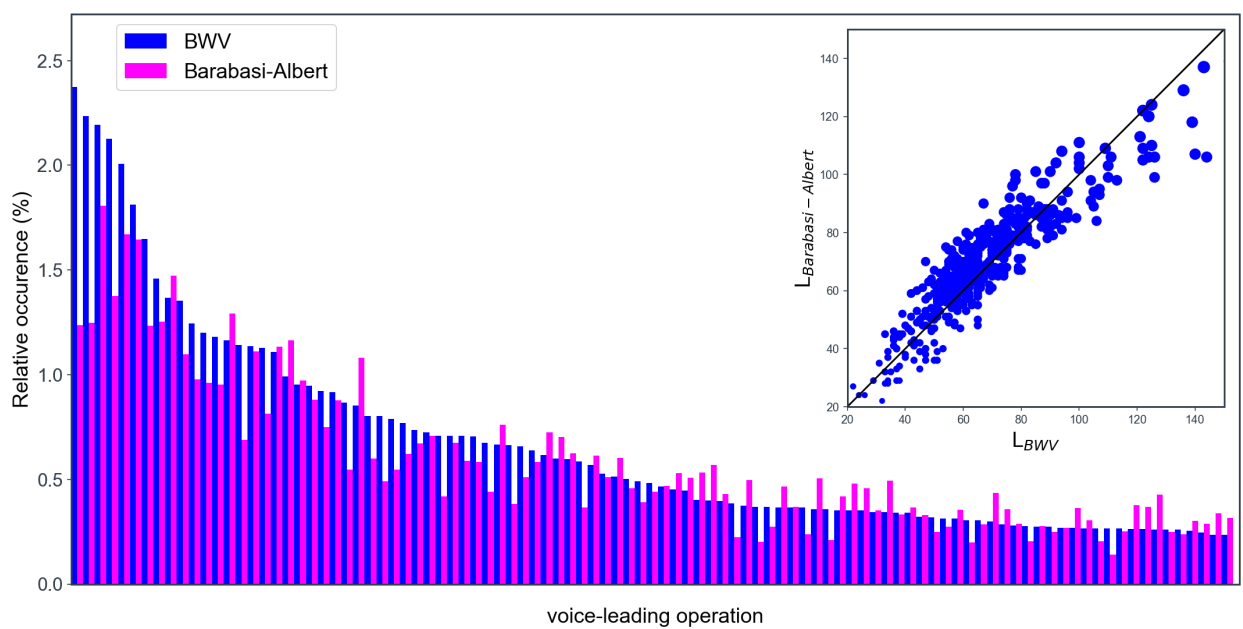

**Figure 10**. Main panel: Relative occurrence of voice-leading operations in the Eulerian paths of the original scores (blue) and in the random scale-free networks generated with the Barabasi-Albert model of preferential attachment (magenta). Inset: comparison between the lengths of the Eulerian circuit for the original ($L_{BWV}$) and the Barabasi-Albert ($L_{Barabasi-Albert}$) network.

To verify the generality of these observations, we performed this procedure for all the scores in the Bach's corpus of `music21` and summarized the results in Figure 10. As in the case of the BWV 267 example, the Eulerian paths on the score networks are systematically shorter than in the original versions due to the omission of double-counting loops in the latter (see Figure S4 in SOM for a comparison of original vs. Eulerian length). However, the length distribution of the Eulerian path on the Barabasi-Albert random graphs is remarkably similar to the one on the original score network, although the topology of the links is now randomly chosen by the preferential attachment algorithm (see inset of Figure 10).

In order to quantify the "tonal character" of the chord sequence corresponding to the Eulerian path on the random scale-free networks generated with the Barabasi-Albert model, we compare in the main panel of Figure 10 the relative occurrence of voice-leading operators, *i.e.* the relative frequency of specific chord progressions, in both the Eulerian path of the original score network and the Barabasi-Albert graph (for a definition of voice-leading operators see Section S2 in the SOM and (Buongiorno Nardelli, Topology of Networks in Generalized Musical Spaces, 2020)). The analysis shows a remarkable agreement between the occurrence of specific progressions in the two systems. In both cases, the most dominant voice-leading operation correspond to the most common progressions of tonal harmony: V-I, and V7-I, followed by I-iii, IV-vi, I-V, IV-I, ii-IV etc. Indeed, irrespective of the completely different connectivity of the two networks, the very geometrical characteristics of the optimal path point to the origin of the *directedness* and *centrality* in tonal compositions as constraints that emerge naturally from the network topology. In particular, the presence of chord "hubs" forces the path to return to these nodes in order to complete the circuit, thus naturally enforcing the existence of harmonic centres. From here, one can envision a compositional design process comprised of the following steps: the generation of a network with the desired topological characteristics, the assignment of a chord distribution to the nodes, and the determination of the optimal path. Of the steps above, the assignment of chords to nodes is definitely a very arbitrary task that can be undertaken in a myriad of ways depending on the choices of the composing agent (being it human or machine). All the characteristics of tonality come into play here: the most obvious, centrality, coincides with the existence of chord hubs, so that highest degree nodes can be assigned to the most prevalent chords (tonic, dominant, etc.) according to one of the numerous fitness models (see for instance (Pardo & Birmingham, 2002), (Navarro-Caceres, Caetano, & Bernardes, 2020)), functionality scores (Cuthbert, 2010) in the preferential attachment, according to degree distributions in existing corpora (Moss, Neuwirth, Harasim, & Rohrmeier, 2019) or using the relative occurrences in chord progressions, as suggested by the results of this study. However, ultimately this is an aesthetic decision that must be undertaken by the composer.

## Conclusions

In conclusion, with this study we have built on the concept of network representation of musical spaces and introduced the idea of a composition as a dynamical score network, we have developed an efficient algorithm for the identification of tonal regions that relies only on the topology of the network, we characterized the full chord progression of the composition as a route optimization problem and shown how these principles can be used to devise a generative model of tonal compositional design.

## Acknowledgements

I wish to thank Dmitry Tymoczko and Mark Newman for long and stimulating discussions, the reviewers of this paper and Ian Quinn for their careful and detailed comments and suggestions.

I am particularly grateful for the support of the University of North Texas and my colleagues of the College of Music, Joe Klein, David Bard-Schwarz and David Stout. Some of this work has been motivated by engaging conversations at the Santa Fe Institute, in particular with Jennifer Dunne, Stefani Crabtree, Miguel Fuentes and Chris Kempes. I will also be forever grateful to IMéRA, the Institute for Advanced Studies of Aix-Marseille University, of which I am proud to be Associate Fellow and where these ideas have been developed in the Spring of 2019 during a sabbatical residency. Finally, special thanks to Richard Kronland, Pascale Hurtado, Sølvi Ystad, Jacques Sapiega, and Mitsuko Aramaki for their friendship and support and Bruno Buongiorno Nardelli for useful discussions and pythonic exchanges.

## Disclosure statement

No potential conflict of interest was reported by the author.

# Bibliography

Albert, R., & Barabási, A. (2002). Statistical mechanics of complex networks. *Reviews of Modern Physics, 74*(1), 47–97.

Alstott, J., Bullmore, E., & Plenz, D. (2014). powerlaw: a Python package for analysis of heavy-tailed distributions. *PLoS ONE, 9*(1), e85777.

Barabasi, A., & Albert, R. (1999). Emergence of scaling in random networks. *Science, 286*, 509-512.

Barabasi, A.-L., & Posfai, M. (2016). *Network Science.* Cambridge: Cambridge University Press.

Bunke, H., & Shearer, K. (1998). A graph distance metric based on the maximal common subgraph. *Pattern Recognition Letters, 19*(3-4), 255-259.

Buongiorno Nardelli, M. (2019). MUSICNTWRK: data tools for music theory, analysis and composition,. *Computer Music Multidisciplinary Research.* Marseille: Springer Lecture Notes in Computer Science.

Buongiorno Nardelli, M. (2020). Topology of Networks in Generalized Musical Spaces. *Leonardo Music Journal, 30*, 38-43.

Callender, C., Quinn, I., & Tymoczko, D. (2008). Generalized Voice-Leading Spaces. *Science, 320*, 346.

Chew, E. (2014). *Mathematical and Computational Modeling of Tonality.* Springer US .

Cuthbert, M. S. (2010). "Music21: A Toolkit for Computer-Aided Musicology and Symbolic Music Data. *International Society for Music Information Retrieval*, (pp. 637-642).

Edmonds, J., & Johnson, E. (1973). Matching, Euler tours and the Chinese postman. *Mathematical Programming, 5*(1), 111-114.

Euler, L. (1739). *Tentamen novae theoriae musicae ex certissismis harmoniae principiis dilucide expositae.* Saint Petersburg Academy.

Fryzlewicz, P. (2014). Wild binary segmentation for multiple change-point detection. *The Annals of Statistics, 42*(6), 2243–2281.

Hagberg, E. A., Schult, D. A., & Swart, P. J. (2008). Exploring network structure, dynamics, and function using NetworkX. In T. Vaught, & J. Millman (Ed.), *Proceedings of the 7th Python in Science Conference (SciPy2008).* Pasadena, CA.

Hedges, T., & Rohrmeier, M. (2011). Exploring Rameau and Beyond: A Corpus Study of Root Progression Theories. In C. Agon, M. Andreatta, G. Assayag, J. Bresson, & J. Manderau, *Mathematics and Computation in Music. Lecture Notes in Artificial Intelligence (6726)* (pp. 334-337). Berlin: Springer.

Heinichen, J. D. (1969). Der General-Bass in der Composition.

Lerdahl, F., & Jackendoff, R. S. (1983). *A Generative Theory of Tonal Music.* Cambridge: MIT Press.

Micchi, G., Gotham, M., & Giraud, M. (2020). Not All Roads Lead to Rome: Pitch Representation and Model Architecture for Automatic Harmonic Analysis. *Transactions of the International Society for Music Information Retrieval, 3*(1), 42-54.

Moss, F. C., Neuwirth, M., Harasim, D., & Rohrmeier, M. (2019). Statistical characteristics of tonal harmony: A corpus study of Beethoven's string quartets. *PLOS One*, 1-16.

Navarro-Caceres, M., Caetano, M., & Bernardes, G. (2020). Objective Evaluation of Tonal Fitness for Chord Progressions Using the Tonal Interval Space. In J. Romero, E. Aniko, & J. Correia, *Artificial Intelligence in Music, Sound, Art and Design* (pp. 150-164). Springer International Publishing.

Neuwirth, M., Harasim, D., Moss, F. C., & Rohrmeier, M. (2018). The Annotated Beethoven Corpus (ABC): A Dataset of Harmonic Analyses of All Beethoven String Quartets. *Frontiers Dig. Human., 5*(16).

Pardo, B., & Birmingham, W. P. (2002). Algorithms for chordal analysis. *Computer Music Journal, 26*(2), 27-49.

Quinn, I. (2010). Are Pitch-Class Profiles Really "Key for Key"? *Zeitschrift der Gesellschaft für Musiktheorie, 7/2*, 151-163.

Rorhmeier, M. (2011). Towards a Generative Syntax of Tonal Harmony. *J. Mathematics & Music, 5*(1), 35-53.

Schenker, H. (1954). *Harmony.* Cambridge: MIT Press.

Schoenberg, A. (1969). *Structural Functions of Harmony.* W. W. Norton & Company.

Shavin, D. (2000). ThinkingThrough Singing. *Fidelio, 9*(4), 60-84.

Temperley, D. (1999). What's Key for Key? The Krumhansl-Schmuckler Key-Finding Algorithm Reconsidered. *Music Perception, 17*(1), 65-100.

Truong, C., Oudre, L., & Vayatis, N. (2020). Selective review of offline change point detection methods. *Signal Processing, 167*, 107299.

Tymoczko, D. (2003). Root Motion, Function, Scale-degree: A Grammar for Elementary Tonal Harmony. *Musurgia, X*(304), 35-64.

Tymoczko, D. (2006). The Geometry of Musical Chords. *Science, 313*, 72-75.

Tymoczko, D. (2012). The Generalized Tonnetz. *Journal of Music Theory, 56*(1), 1-52.

Tymoczko, D., Gotham, M., Cuthbert, M. S., & Aritza, C. (2019). The Romantext Format: A flexible and standard method for representing Roman numeral analyses. *International Society for Music Information Retrieval Conference (ISMIR 2019).* Delft.

Zinoviev, D. (2018). *Complex Network Analysis in Python: Recognize - Construct - Visualize - Analyze – Interpret.* Pragmatic Bookshelf.

# Tonal harmony and the topology of dynamical score networks

## Supplementary Online Material

Marco Buongiorno Nardelli

*Department of Physics and Division of Composition Studies, University of North Texas, Denton, TX, 76203, USA*

*CEMI – Center for Experimental Music and Intermedia, University of North Texas, Denton, TX, 76203, USA.*

*iARTA – Initiative for Advanced Research in Technology and the Arts, University of North Texas, Denton, TX, 76203, USA.*

E-mail: mbn@unt.edu, websites: www.materialssoundmusic.com, www.musicntwrk.com

Tel. +1 (940) 369-8596

This pdf contains:

1. Supplemental Section S1. Score filtering procedure including Figures S1 and S2
2. Table S1.
3. Supplemental Figure S3
4. Supplemental Figure S4
5. Supplemental Section S2. Voice-leading operators.
6. Bibliography cited.

**Supplemental Section S1. Score filtering procedure**

      To illustrate this initial data preparation process we will consider the first movement of the Beethoven string quartet Op. 127 and then extend the analysis to the full piece. We use the score from the Annotated Beethoven Corpus (ABC-1.0) (Neuwirth, Harasim, Moss, & Rohrmeier, 2018) exported to musicxml using the MuseScore scorewriter app (at www.musescore.com). The first movement of the quartet is comprised of 282 bars, for a total of 1298 individual pcsets in the sequence. Of these 1298, 246 are the unique pcs combinations that are used throughout. If we number individually each unique pcset, we can build a map of the occurrences of the various chords and construct a time-series as the one displayed in the main panel of Figure S1. We exploit the above observations to curate the data by filtering out all the pcsets that occur a number of times that lies below a certain threshold. From the analysis of the histogram of occurrences (shown in the inset of Figure S1), we observe that there is a clear separation between a few pcsets that have high occurrences, and the rest of them with lower occurrences, many of which appear only once in the course of the piece. Note that here we do not make any assumption on which pcset is more "important" although, as we argue in the main paper, this is a clear quantitative manifestation of centrality in tonal music. With this observation in mind, we can now decide a filtering threshold and generate the set of data for further analysis. We choose to fix the threshold for filtering pcsets to 10%, that is we ignore all pcsets that have occurrences less that 10% of the max (here the reference is pcset n. 0 with 109 occurrences and that happens to be the Eb major triad [7,10,2], the key of the composition). Applying this procedure, we generate the "filtered" sequence of pcsets displayed in Figure S3 (and also displayed in Figure 2 in the main paper). Now, the number of time stamps (the successive occurrences of pcsets) is reduced to 752, with only 33 unique chords. We are now ready to use these data for a more complete analysis of the score network.

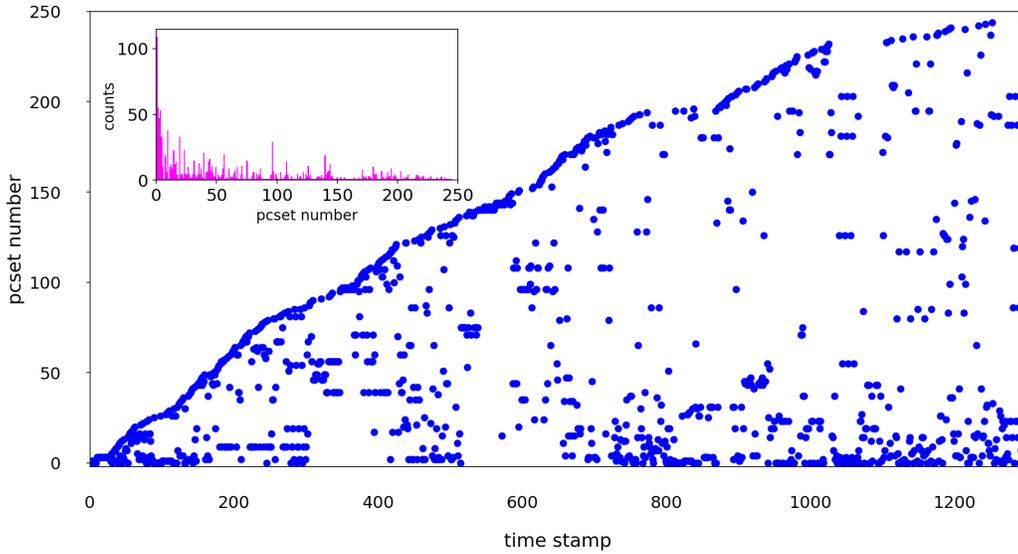

**Figure S1.** Time series of the full sequence of all the normal-ordered pcsets extracted from the score of the first movement of the string quartet Op. 127 n. 12. Inset: Histogram of occurrences of each individual pcset. We plot here the number of times a given pcset appears in the composition. Again, we see a predominance of a few chords with high occurrence, another evidence of the centrality property.

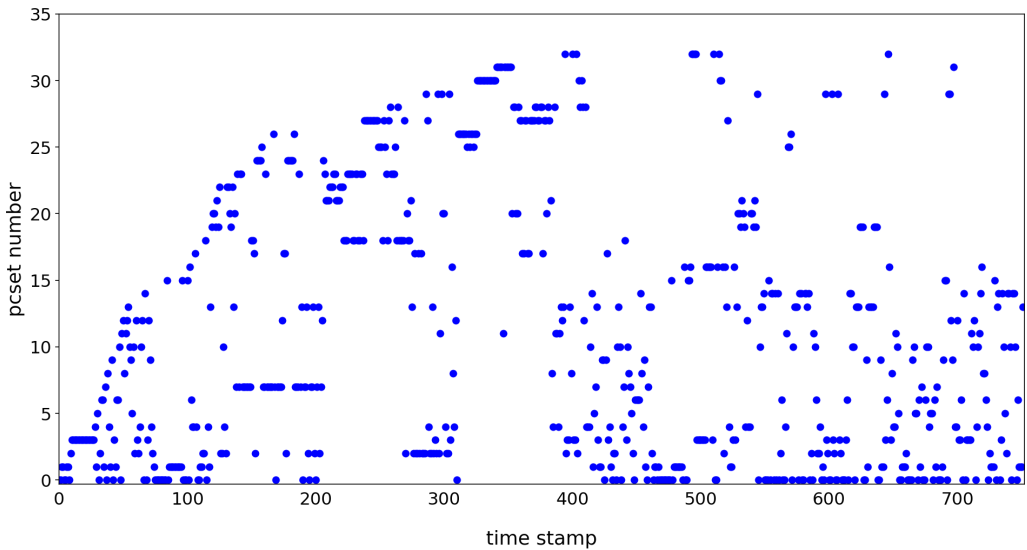

**Figure S2**. Time series of the filtered sequence of the normal ordered pcsets from the first movement of the string quartet Op. 127 n. 12.

**Table S1**. Harmonic region assignments in the ABC compared to the automatic segmentation of the dynamical score network (this work). The notation is the same as in the ABC.

mov. 1

| bar | region (ABC) | region (this work) |
|---|---|---|
| 1-34 | Eb:I | Eb:I |
| 34-36 | I | iii |
| 36-39 | vi | iii |
| 39-63 | iii | iii |
| 63-75 | iii | III |
| 75-101 | III | III |
| 101-107 | III | vi |
| 107-117 | vi | vi |
| 117-124 | I | vi |
| 124-129 | I | ii |
| 129-133 | ii | ii |
| 133-135 | ii | VI |
| 135-147 | VI | VI |
| 147-150 | I | VI |
| 150-162 | I | I |
| 162-166 | ii | VI |
| 166-167 | ii | I |
| 167-187 | I | I |
| 187-199 | IV | IV |
| 199-203 | ii | IV |
| 203-207 | I | IV |
| 207-282 | I | I |

mov 2.

| bar | region (ABC) | region (this work) |
|---|---|---|
| 1-4 | Ab:I | Ab:V |
| 4-27 | I | I |
| 27-29 | I | vi |
| 29-32 | I | V |
| 32-37 | I | I |
| 37-38 | IV | I |
| 38-48 | I | I |
| 48-50 | V | I |
| 50-61 | I | I |
| 62-62 | I | #V |
| 62-81 | #V | #V |
| 81-87 | I | I |
| 87-88 | V | I |
| 88-99 | I | I |
| 99-104 | IV | IV |
| 104-110 | #iii | #iii |
| 110-111 | #I | #iii |
| 111-130 | I | I |

mov. 3

| bar | region (ABC) | region (this work) |
|---|---|---|
| 1-8 | Eb:I | Eb:I |
| 8-28 | I | I |
| 28-31 | I | V |
| 31-35 | V | V |
| 35-38 | V | @none |
| 38-41 | @none | @none |
| 41-43 | vi | @none |
| 43-51 | vi | vi |
| 51-57 | vi | iv |
| 57-59 | i | iv |
| 59-60 | i | bIII |
| 61-66 | bIII | bIII |
| 66-67 | iii | bIII |

mov. 4

| bar | region (ABC) | region (this work) |
|---|---|---|
| 1-22 | Eb:I | Eb:I |
| 22-44 | I | V |
| 44-52 | V | V |
| 52-53 | vi | V |
| 53-67 | V | V |
| 67-73 | @none | @none |
| 73-80 | V | II |
| 80-90 | V | I |
| 90-105 | I | I |
| 105-106 | I | ii |
| 106-111 | VI | ii |
| 111-126 | vi | vi |
| 126-131 | ii | ii |

| | | | | | | |
|---|---|---|---|---|---|---|
| 67-79 | iii | iii | | 131-135 | II | ii |
| 79-81 | vi | iii | | 135-139 | I | ii |
| 81-88 | ii | ii | | 139-140 | I | I |
| 88-89 | ii | I | | 140-145 | IV | I |
| 89-111 | I | I | | 145-170 | IV | IV |
| 111-113 | vi | I | | 170-173 | I | IV |
| 113-145 | I | I | | 173-184 | IV | IV |
| 145-146 | I | i | | 184-255 | I | I |
| 146-153 | i | i | | 255-256 | IV | I |
| 153-163 | i | bVII | | 256-261 | IV | IV |
| 163-188 | bVII | bVII | | 261-262 | vi | vi |
| 188-189 | bVII | v | | 262-263 | IV | IV |
| 189-204 | v | v | | 263-264 | IV | I |
| 204-210 | v | bIII | | 264-267 | #iv | I |
| 210-223 | bIII | bIII | | 267-299 | I | I |
| 223-224 | i | bIII | | | | |
| 224-238 | i | i | | | | |
| 238-246 | i | V | | | | |
| 246-273 | V | V | | | | |
| 273-292 | I | I | | | | |

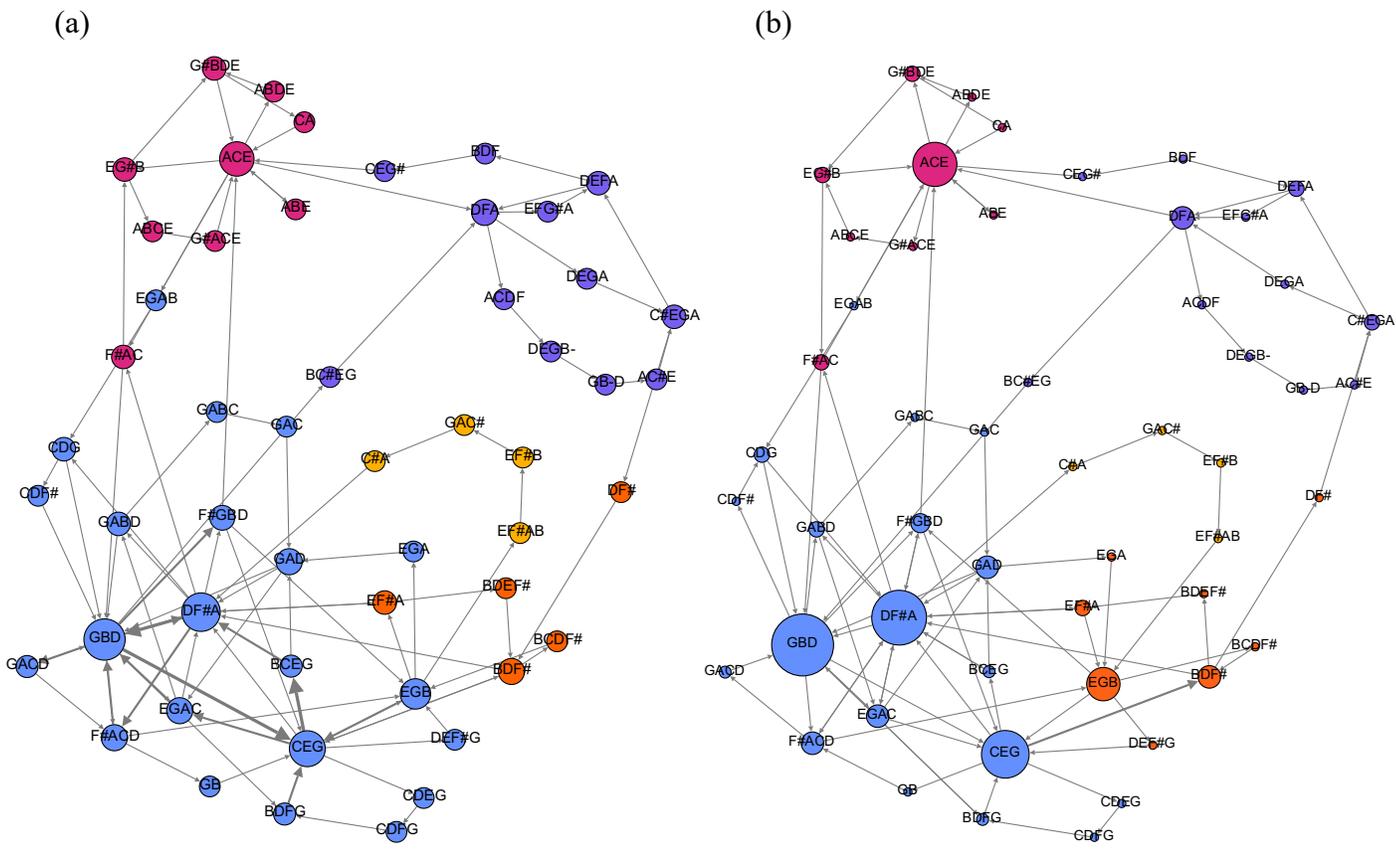

**Figure S3.** (a) Score network of BWV 267 (same as Figure 1 in main paper) with the original directional edges (chord progressions), degrees and modularity classes; (b) same network as (a) but with the directional edge distribution from a minimal Eulerian path. The behavior of degree distribution and modularity classes reflect the reduction in path length, but trends largely agree with the original version.

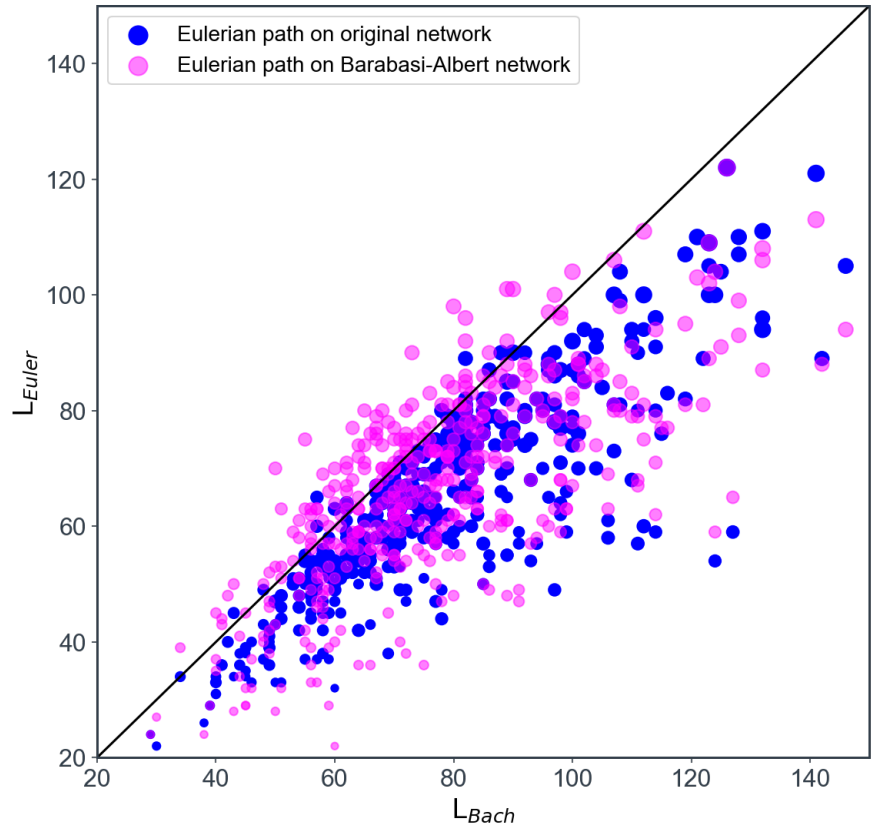

**Figure S4.** Comparison between the length of the original chorale score (# of pcsets) *vs.* the length of the Eulerian path on the same score network (blue dots) or on a Barabasi-Albert network (magenta dots) with matching degree node assignment (see main paper for discussion).

**Supplemental Section S2. Voice-leading operators**

In (Buongiorno Nardelli, 2020) we have introduced a quantification of voice leading through distance operators. For this we use the minimal Euclidean voice leading distance for arbitrary TET-notes temperaments (12, 24, etc.) defined as:

$$d_{\min}(\mathbf{x},\mathbf{y}) = \min_{\mathbf{TET}_j} \sqrt{\sum_i \left(\vec{x}_i - (\vec{y}_i \pm \mathbf{TET}_j)\right)^2}. \tag{S1}$$

Here $\vec{x}_i$ and $\vec{y}_i$ are the pcsets in ascending order and $\mathbf{TET}_j = (0,0,\ldots,0,\pm\text{TET},0,\ldots)$, a vector of dimension $N_C$ that raises or lowers the $j^{\text{th}}$ pitch of the ordered pcset by TET. It is easy to verify that this definition of distance operator is equivalent to finding the minimal distance between all possible cyclic permutations of the pcset. Such definition is easily extended to non-bijective voice-leadings, by an iterative duplication of pitches in the smaller cardinality pcset, and then looking for the multiset that produces the minimal distance.

With this definition, we can introduce "distance" operators, $\mathbf{O}(\{n_i\})$, which raise or lower by an integer $n$ the $i^{\text{th}}$ component of a vector of dimension $N_C$. With this definition, if $x$ is transformed into $y$ by $\mathbf{O}(\{n_i\})$, then:

$$d(\mathbf{x},\mathbf{y}) = \sqrt{\sum_i n_i^2}. \tag{S2}$$

With this definition, the distance operator is assumed to be raising or lowering by the specified amount one of the pitches of the pcset with no information on the positional ordering ($i$ is left unspecified). For instance, as distance operator, $\mathbf{O}(1)$ applied to the [0,4,7] pcset (C Maj chord) generates the following chords: [0, 3, 7] (C min), [0, 4, 6] (C incomplete half-diminished seventh), [0, 4, 8] (C augmented), [5, 7, 0] (F quartal trichord), [1, 4, 7] (C# diminished), [4, 7, 11] (E min). It is a simple extension to choose a positional specification (specify $i$), in order to obtain a unique (bijective) mapping. If the number of specified components is smaller than the cardinality of the pcset, all unspecified components are assumed to be 0.

A generalization of distance operators leads to the definition of normal-ordered voice leading operators $\mathbf{O}_{\text{VL}}(\{n_i\})$, that given a normal-ordered pcset, transform it into the successive normal-ordered pcset in the chord progression. Here $\{n_i\}$ is an ordered vector of positive or negative integers where each component represents the minimal number of steps that need to be applied to the corresponding pitches of the ordered pcset. So, for instance, $\mathbf{O}_{\text{VL}}(-1,-2,0)$ applied to [0,4,7] produces [11,2,7] ([7,2,11] when brought back to normal order), the I-V progression from the C major triad to its dominant in the key of C.

# Bibliography

Buongiorno Nardelli, M. (2020). Topology of Networks in Generalized Musical Spaces. *Leonardo Music Journal, 30*, 38-43.

Neuwirth, M., Harasim, D., Moss, F. C., & Rohrmeier, M. (2018). The Annotated Beethoven Corpus (ABC): A Dataset of Harmonic Analyses of All Beethoven String Quartets. *Frontiers Dig. Human., 5*(16).